\documentstyle[preprint,aps,epsf]{revtex}
\begin{document}
\preprint{DOE/ER/40427-25-N93}
\draft
\title{ELECTROMAGNETIC GAUGE INVARIANCE OF\\
CHIRAL HYBRID QUARK MODELS}
\author{W. Koepf and E.M. Henley}
\address{
Department of Physics, FM-15, University of Washington,
Seattle, WA 98195}
\date{\today}
\maketitle
\begin{abstract}
In this work, we investigate the question, whether the conventional
analysis of the electromagnetic form factors of the nucleon,
evaluated in the framework of the cloudy bag model (CBM) or other
chirally invariant hybrid quark models utilizing the same philosophy,
is gauge invariant.  In order to address that point, we first
formulate the CBM in a style that resembles the technique of loop
integrals.  Evaluating the self energy and the electromagnetic form
factors of the nucleon in that manner and comparing with the standard
analysis where nonrelativistic perturbation theory is used, allows us
to show that our approach is appropriate and to point out, what
approximations are made in the standard derivation of the model.
{}From the form of those loop integrals, we then show that additional
diagrams are needed to preserve electromagnetic gauge invariance and
we assess the corresponding corrections.
\end{abstract}

\bigskip
\newpage
\section{INTRODUCTION}

The MIT bag model [1] provides the simplest relativistic description
of the nucleon's quark substructure and it yields confinement as well
as a reasonable description of the standard nucleonic properties.
In its original form, however, the model violates chiral symmetry, an
important symmetry of the strong interactions [2].  But it is well
known, that by introducing a suitable interaction of the quarks with
an elementary pion field, manifest chiral symmetry can be
restored [3].

In the cloudy bag model (CBM) of Miller et al. [4,5,6], the exterior
pionic field is quantized and it penetrates the quark bag.  While the
corresponding Lagrangian is highly non-linear, practical calculations
are carried out as a perturbative expansion -- in $1/f_\pi$ -- where
only the leading order terms are kept.  The model yields quite
satisfactory results for many hadronic properties such as charge
radii and magnetic moments [5,6].  Following the approach which was
used to obtain the CBM, Williams and Dodd [7] later introduced other
hybrid quark models and investigated chiral extensions of both the
Nielsen-Patkos color dielectric model [8] and the Friedberg-Lee
soliton model [9].

In a recent publication by Musolf et al. [10], it was argued that the
standard analysis of the electromagnetic form factors of the nucleon
in the framework of the cloudy bag model is not gauge invariant.

Here, we investigate this issue by formulating the CBM in a way
that resembles the technique of loop integrals. We first evaluate
the nucleon's self energy and its electromagnetic Sachs form factors
[11], $G_E$ and $G_M$, and compare with the standard analysis which
uses nonrelativistic perturbation theory. By showing that both
results agree, we will prove that our approach is appropriate and
this will enable us to point out what approximations are employed
in the usual derivation of the CBM.

We conclude from this analysis that the standard treatment of
the CBM is not gauge invariant, because it involves an explicitly
momentum dependent meson-nucleon vertex. We then derive a ``seagull"
(contact) correction either by following Gross and Riska [12] or by
applying the minimal substitution prescription as outlined by Ohta
[13], and discuss the corresponding alterations.

The outline of this work is as follows.  In Sect. 2, we give a brief
introduction to the physics of the cloudy bag model; in Sect. 3 we
describe the construction of ``physical nucleons" in the framework
of the model and evaluate their self energy. In Sect. 4, we study
the coupling of these ``physical nucleons" to an external photon; in
Sect. 5 we discuss the $U(1)$ gauge invariance of that coupling and
derive ``seagull" corrections. A few concluding remarks are made in
Sect. 6.

\section{THE CLOUDY BAG MODEL}

The very basis of the CBM -- and also of other chirally invariant
extensions of effective soliton models, as introduced, for example,
by Williams et al. [7] -- is to supplement a simple model which
describes quark degrees of freedom, i.e. the MIT bag model [1] or the
Friedberg-Lee soliton model [9], with a non-linear realization of
chiral symmetry.  The corresponding Lagrangian for the CBM reads,
after the usual linearization and if only the leading order terms are
kept [4],
\begin{equation}
{\cal L}~ = ~{\cal L}_{MIT}+{\cal L}{\pi}+{\cal L}_{IA}^{S,V} \ ,
\end{equation}
with
\begin{eqnarray}
{\cal L}_{MIT} & = & \bigl[ \, \bar q \left( i\gamma^\mu \partial_\mu
- m_q \right) q ~-~B \, \bigr] \, \Theta_V ~-~
{\scriptstyle{1\over 2}} \bar q q \Delta_S \ , \\
{\cal L}{\pi} & = & {\scriptstyle{1\over 2}} \left( \partial_\mu
\mbox{\boldmath $\pi$} \cdot \partial^\mu \mbox{\boldmath $\pi$}
{}~-~m_\pi^2 \mbox{\boldmath $\pi$}^2 \right) \ ,
\end{eqnarray}
\begin{mathletters}
\begin{eqnarray}
{\cal L}_{IA}^S &  = - & {i \over 2 f_\pi} \, \bar q \, \gamma_5 \,
\mbox{\boldmath $\tau$} \cdot \mbox{\boldmath $\pi$} \, q \, \Delta_S
\ , \\
{\cal L}_{IA}^V~ & = &   {1 \over 2 f_\pi} \, \bar q \, \gamma^\mu \,
\gamma_5 \, \mbox{\boldmath $\tau$} \cdot \partial_\mu
\mbox{\boldmath $\pi$} \, q \, \Theta_V \ .
\end{eqnarray}
\end{mathletters}
Here, $\Theta_V$ is a step-function, which is 1 on the inside of the
bag and 0 on the outside, and the surface delta function $\Delta_S$
is its derivative. The interaction term ${\cal L}_{IA}^S$ of Eq. (4a)
refers to pseudoscalar coupling (the surface coupled version of the
CBM [4]) and ${\cal L}_{IA}^V$ of Eq. (4b) refers to pseudovector
coupling (the volume coupled version [14]).

${\cal L}$, which is a functional of the elementary quark and meson
fields, is invariant under global $U(1)$ gauge transformations and
the corresponding conserved current is
\begin{equation}
J^\mu = e\bar q \gamma^\mu Q q \Theta_V~-~
ie\left(\phi^+ \partial^\mu \phi ~-~ \phi \partial^\mu \phi^+ \right)
{}~+~ {ie \over 2 f_\pi} \bar q \gamma^\mu\gamma_5
\left(\tau_- \phi~+~\tau_+ \phi^+\right) q \Theta_V \ ,
\end{equation}
where $\phi=(\pi_1+i\pi_2)/\sqrt{2}$ destroys a $\pi^-$, with
$\tau_\pm=\mp(\tau_1\pm i \tau_2)/\sqrt{2}$, and where $Q$ is the
quark charge matrix. The last term, which corresponds to a contact
diagram, arises for pseudovector coupling only. At this stage, the
relevant degrees of freedom are quarks and mesons and there is
obviously no problem with the electromagnetic $U(1)$ gauge
invariance.

Using the ground state wave function of the MIT-bag [1],
\begin{equation}
\psi_s ({\bf r}, t)~=~{\cal N}~\left(\begin{array}{c}
j_0(\omega r/R) \\ i \mbox{\boldmath $\sigma$} \cdot \hat{\bf r} \,
j_1(\omega r/R) \end{array} \right)
e^{-i\omega t/R}~\theta (R-r)~\chi_s^{[1/2]}
\ ,
\end{equation}
where $R$ is the bag radius, $\omega=2.04$ is the dimensionless bag
eigenfrequency and ${\cal N}$ is a normalization factor, and after a
decomposition of the pionic field into plane waves, we find from Eqs.
(4a) or (4b) for the interaction Hamiltonian,
\begin{equation}
H_I~=~-~\int\! d^3{\bf r}~{\cal L}_{IA}^{S,V}
\end{equation}
in second-quantized notation and projected on the space of physical
particle states which are members of the {\bf 56} representation of
$SU(6)$,
\begin{equation}
H_I(A,B)~=~-ig \int\! d^3{\bf k}~<\!B_{s'}|\mbox{\boldmath $\sigma$}
\cdot {\bf k} \, \tau_j|A_s\!>~u(|{\bf k}|R)~B_{s'}^\dagger \, A_s~
a^\dagger_{j,{\bf k}}~+~h.c.
\ .
\end{equation}
Here, $A_s$ destroys a bag $A~\epsilon~\{\bf 56\}$ with spin $s$,
$B_{s'}^\dagger$ creates a bag $B~\epsilon~\{\bf 56\}$ with spin
$s'$, and $a^\dagger_{j,{\bf k}}$ creates a physical on-mass-shell
pion with isospin $j$ and three-momentum ${\bf k}$.  The vertex
function, $u(|{\bf k}|R)$, has the form
\begin{equation}
u(|{\bf k}|R)~=~{3 j_1(|{\bf k}|R) \over |{\bf k}|R} \ ,
\end{equation}
and for the coupling constant we find
\begin{equation}
g~=~{\omega \over 6 f_\pi (\omega-1)}~=~{3 \sqrt{4\pi} f_{\pi AB}
\over 5 m_\pi} \ ,
\end{equation}
which defines the pseudoscalar pion-nucleon coupling, $f_{\pi NN}$.
Usually [4], one replaces the microscopically determined quantity
($f_{\pi NN}=0.23$) with the experimental value ($f_{\pi NN}=0.28$).
The other $f_{\pi NB}$ are then given by spin-isospin $SU(6)$.

The above result holds for pseudoscalar coupling (see Eq. (4a)) as
well as for pseudovector coupling (see Eq. (4b)). The vertex function
in Eq. (9) resembles the final extension of the baryonic bag and
its internal structure.  It is that explicitly momentum dependent
vertex which spoils gauge invariance, as we will see in the
following.
When applying this formalism to other chirally invariant hybrid quark
models [7], the only modification is that, due to different quark
wave functions, another vertex function will appear. Aside from that,
the rest of this investigation will apply.

\section{PHYSICAL BARYONS}

In all hybrid quark models, the physical baryons are ``dressed bags".
This means, that because of the coupling of the pion field to the
quarks in the bag, the ``physical nucleon", $|N\!>$, will be part of
the time a bare three-quark nucleonic MIT bag, $|N_0\!>$, and part of
the time a baryon, $|B_0\!>\epsilon~\{\bf 56\}$, with one pion ``in
the air" [6]:
\begin{equation}
|N\!>~=~Z^{-1}~\left(|N_0\!>~+~\sum_{B_0}|B_0,\pi\!>\right) .
\end{equation}
Here, $Z$ is a wave function renormalization factor which guarantees
that $<\!N|N\!>=1$. All other possible configurations would have more
than one pion present at any time and are neglected with the argument
that they are suppressed through larger energy denominators [4].

Using the pseudoscalar meson-nucleon coupling, the pionic
contribution to the nucleon's self energy can then be written as
[15],
\begin{equation}
\Sigma (p)~=~ig^2\int\!{d^4k \over (2\pi)^4}~\gamma_5~f(k)~
S(p - k)~f(k)~\gamma_5~\Delta(k) \ ,
\end{equation}
where $f(k)=u(|{\bf k}|R)$ is the momentum dependent vertex function
of Eq. (9), $g$ is the coupling strength from Eq. (10), which also
implies a sum over all possible intermediate spin and isospin states,
and with the free fermionic and mesonic propagators, $S(p-k)$ and
$\Delta(k)$, to be defined shortly.

If we employ a nonrelativistic approximation and also neglect the
recoil which the nucleon experiences when it emits or absorbs a pion,
we find the replacement,
\begin{equation}
S(p-k)~\equiv~{1 \over \rlap{/} p - \rlap{/} k - m_B} ~\to~
{1 \over E - k_0 - m_B} \ ,
\end{equation}
where $E$ is the zeroth component of the nucleon's four momentum
$p=(E,{\bf 0})$ and where $m_B$ is the mass of the intermediate
baryon. If we further place the pion on its mass shell, we find for
its propagator [16],
\begin{equation}
\Delta(k)~\equiv~{1 \over k^2 - m_\pi^2} ~\to~
{-i\pi \over \omega_{\bf k}}~\delta(k_0 - \omega_{\bf k}) \ ,
\end{equation}
with $\omega_{\bf k}=\sqrt{|{\bf k}|^2 + m_\pi^2}$. This finally
yields
\begin{equation}
\Sigma(E)~=~\sum_B~{f_{\pi NB}^2 \over \pi m_\pi^2} \int\! d|{\bf k}|
{|{\bf k}|^4 u(|{\bf k}|R)^2 \over \omega_{\bf k}
(E-\omega_{\bf k}-m_B)} \ ,
\end{equation}
which agrees with Eq. (3.4c) of ref. [5]. From this expression we can
determine the wave function renormalization factor, $Z$, using [5]
\begin{equation}
Z~\equiv 1-{\partial \Sigma (E) \over \partial E}
\Bigg\bracevert_{E=m} =1 + \sum_B Z_B = 1 +\sum_B{f^2_{\pi NB}\over
\pi m^2_\pi} \int d|{\bf k}|
{|{\bf k}|^4 u(|{\bf k}|R)^2 \over \omega_{\bf k} (\omega_{\bf
k}+\omega_B)^2} \ ,
\end{equation}
where $\omega_B=m_B-m$ and $m$ is the nucleon mass. $Z^{-1}$ is the
probability that the ``physical" nucleon is a bare three-quark
state [5].

\section{THE ELECTROMAGNETIC CURRENT}

In the following, we will evaluate matrix-elements of the
electromagnetic current, $J^\mu$ of Eq. (5), sandwiched between the
``physical nucleon" states as defined in Eq. (11). Thereby, we will
neglect all internal excitations of the bag and also its recoil.
With these assumptions, our investigation is limited to the Breit
frame [17], where the energy of the exchanged virtual photon
vanishes. Furthermore, the bag states we are considering are
localized and are therefore not eigenstates of the momentum operator.
This rather crude approximation can be cured somewhat by center of
mass corrections to the electromagnetic observables [18] or, more
satisfactorily, by boosting the bag [19]. The modifications which
arise from this procedure are, however, not the subject of this
investigation.

The expression for the electromagnetic (e.m.) vertex function of
the coupling of an external photon to the pion in the loop [15],
again using the pseudoscalar meson-nucleon coupling, is
\begin{equation}
\Gamma^\mu_M (q^2)~=~ig^2e_M\int\! {d^4k \over (2\pi)^4}~\gamma_5~
f(k)~ S(p-k)~f(k')~\gamma_5~\Delta(k)~(k+k')^\mu~\Delta(k') \ ,
\end{equation}
with $k'=k+q$ and where $e_M$ is the charge of the meson. Evaluating
this with the states which were defined in Eq. (11), and performing
the replacements which were described in the last section, we finally
find for the mesonic contribution to the nucleon's electric form
factor
\begin{equation}
G_E^{mes}(|{\bf q}|)={1\over Z}
\sum_{B}{e_Mf_{\pi NB}^2 \over 2\pi^2 m_\pi^2} \int\! d^3{\bf k}
{u(|{\bf k}|R) u(|{\bf k'}|R)~{\bf k} \cdot {\bf k'} \over
(\omega_{\bf k}+\omega_B) (\omega_{\bf k'}+\omega_B)
(\omega_{\bf k}+\omega_{\bf k'})} \ ,
\end{equation}
and the nucleon's magnetic form factor
\begin{equation}
G_M^{mes}(|{\bf q}|)={1\over Z}
\sum_{B}{e_Ms(B)f_{\pi NB}^2 \over 4\pi^2 m_\pi^2}\int\! d^3{\bf k}
{u(|{\bf k}|R) u(|{\bf k'}|R) (\omega_{\bf k}+\omega_{\bf k'}+
\omega_B) |{\bf k} \times \hat{\bf q}|^2 \over
\omega_{\bf k} (\omega_{\bf k}+\omega_B) (\omega_{\bf k}+
\omega_{\bf k'}) \omega_{\bf k'} (\omega_{\bf k'}+\omega_B)} \ ,
\end{equation}
where ${\bf k'} = {\bf k} + {\bf q}$, and $s(B)=1$ if $B$ is an octet
state and $s(B)=-{1 \over 2}$ for a decuplet state. $Z^{-1}$ is the
wave function renormalization factor defined in Eq. (16). Eqs. (18)
and (19) agree with the expressions (4.17), (4.26) and (4.27) of
ref. [5].

By employing the pseudoscalar meson-nucleon coupling, the e.m.
vertex function of the interaction of the external photon with the
initial or the intermediate baryon can be written as [15]
\begin{equation}
\Gamma^\mu_F (q^2)~=~e_N~\gamma^\mu~+~ig^2e_B\int\! {d^4k \over
(2\pi)^4} \gamma_5\;f(k) S(p-k)\gamma^\mu S(p'-k)
f(k)\gamma_5\Delta(k)\ ,
\end{equation}
where $p'=p+q$, $e_N$ is the initial nucleon's charge and $e_B$ is
the charge of the intermediate baryon. This yields
\begin{eqnarray}
G_{E}^{bag}(|{\bf q}|) & = & {1\over Z} G^0_E (|{\bf q}|)
\left(~e_N~+~\sum_B e_B~Z_B \right) \ , \\
G_{M}^{bag}(|{\bf q}|) & = & {1\over Z}~G_{M}^0(|{\bf q}|)~
\left(~\mu_N~+~\sum_{B,B'} \alpha_{B,B'}~\mu_{B,B'}~Z_{B,B'} \right)
\ ,
\end{eqnarray}
with $\mu_p=1$, $\mu_n=-2/3$ and $Z_B$ from Eq. (16). In addition, we
have the $SU(6)$ magnetic moments $\mu_{B,B'}=<\!B|\hat\mu|B'\!>$,
the factors
\begin{equation}
\alpha_{B,B'}~=~\left\{ \begin{array}{ll}
   -1/3 & \mbox{for $B,B'~\epsilon$~{\bf 8}} \nonumber \\
   5/9 & \mbox{for $B,B'~\epsilon$~{\bf 10}} \\
   4/3 & \mbox{for $B~\epsilon$~{\bf 8}, $B'~\epsilon$~{\bf 10}} \ ,
\nonumber \end{array} \right.
\end{equation}
and
\begin{equation}
Z_{B,B'}~=~{f_{\pi NB}f_{\pi NB'} \over \pi m_\pi^2}\int\! d|{\bf k}|
{|{\bf k}|^4 u(|{\bf k}|R)^2 \over \omega_{\bf k} (\omega_{\bf k}+
\omega_B) (\omega_{\bf k}+\omega_B')} \ .
\end{equation}
Here, $G_{E,M}^0$ are the electromagnetic form factors of a bare
MIT-bag,
\begin{mathletters}
\begin{eqnarray}
G_E^0(|{\bf q}|)~& = & 4\pi~{\cal N}^2 R^3 \int_0^1\! dx~x^2~
\left(j_0^2(\omega x)~+~j_1^2(\omega x)\right)~j_0(|{\bf q}|Rx)\ , \\
G_M^0(|{\bf q}|)~& = & 8\pi~e~{\cal N}^2 R^4 \int_0^1\! dx~x^3~
j_0(\omega x)~j_1(\omega x)~{j_1(|{\bf q}|Rx) \over |{\bf q}|Rx} \ ,
\end{eqnarray}
\end{mathletters}
$e$ is the elementary charge and ${\cal N}$ is the normalization
factor which was introduced in Eq. (6).  From Eqs. (21) to (25) we
can derive the baryonic contribution to the magnetic moment and the
charge radius of the nucleon, and the corresponding expressions agree
with the ones given in ref.~[5].

For the pseudovector (volume) coupled version of the CBM, there
exists, in addition to the vertices described above, also a
``seagull" diagram. The resulting loop integral can be written as,
\begin{equation}
\Gamma^\mu_{sg} (q)={i g e_M \over 2 f_\pi}\int\! {d^4k \over
(2\pi)^4} \Bigl(\gamma_5 f(k') S(p-k) \gamma^\mu \gamma_5 \Delta(k')
+ \Delta(k) \gamma^\mu \gamma_5 S(p-k) f(k) \gamma_5 \Bigr)  \ ,
\end{equation}
where $e_M$ is the charge of the meson in the loop. The ``seagull"
graph does not contribute to the electric form factor, as $\bar\psi'
\gamma^0 \gamma_5 \psi$ vanishes, if $\psi'$ and $\psi$ correspond
to the same relativistic orbital wave function. For the respective
magnetic form factor, we find,
\begin{equation}
G_M^{sg}(|{\bf q}|)={\xi\over Z}\sum_{B}
{e_Ms(B)f_{\pi NB}^2 \over 4\pi^2 m_\pi^2}
\!\!\int\!\!{d^3{\bf k}~u(|{\bf k}|R) \over \omega_{\bf k}
(\omega_{\bf k}+\omega_B)}\left(
{{\bf k} \cdot {\bf q} \over |{\bf q}|^2} I_1(|{\bf k'}|R) +
|\hat{\bf k} \times \hat{\bf q}|^2 I_2(|{\bf k'}|R)\right) \ ,
\end{equation}
where $\xi=3/j_0^2(\omega)$ stems from the normalization of the bag
wave function and with
\begin{mathletters}
\begin{eqnarray}
I_1(|{\bf k}|R) & = & \int_0^1\!\!dx\,x^2 \left[
\left( j_0^2(\omega x) - j_1^2(\omega x) \right) j_0(|{\bf k}|R x)~+~
2j_1^2(\omega x){j_1(|{\bf k}|Rx) \over |{\bf k}|Rx} \right] \ , \\
I_2(|{\bf k}|R) & = & \int_0^1\!\!dx\,x^2 j_1^2(\omega x)~
j_2(|{\bf k}|Rx) \ .
\end{eqnarray}
\end{mathletters}
The corresponding magnetic moment, which can be obtained from Eq.
(27), agrees with the expression given in Eq. (4.13) of ref. [14].

\section{GAUGE INVARIANCE}

As we showed in the previous sections, the CBM and other chirally
invariant hybrid quark models can be understood also as effective
field theories with free pions and composite nucleonic fields, the
coupling of which is governed by a momentum dependent vertex
function, $u(|{\bf k}|R)$ of Eq. (9) for the CBM. It has been known
for a long time [20], that some e.m. current flows within the
interaction range delimited by this form factor and that such an
interaction current consequently will modify the electromagnetic
current operator.  As pointed out by Gross and Riska [12] and again
by Musolf et al. [10], if the meson-nucleon coupling is not
pointlike, as, for example, due to the nucleon's internal structure
or finite extension, inclusion of ``seagull" diagrams or some
related procedure (see e.g. ref. [21]) is required to satisfy the
Ward-Takahashi identity [22].  This means, that the presence of the
meson-nucleon form factor will modify the effective current operators
or, alternatively speaking, the momentum dependent vertex of Eq. (9)
will spoil gauge invariance unless a ``seagull" correction is
included.  This correction has not been carried out in most such
investigations [4,5,6,7,23].  Note also that, as shown by Nishijima
[24], the Ward-Takahashi identities are valid -- and have the same
form -- for elementary as well as for composite particles, as they
are a direct manifestation of $U(1)$ gauge invariance.

The corresponding alterations can either be found by following Gross
and Riska [12], where the phenomenological form factor is regarded
as an effective self-energy correction, or by applying the minimal
substitution prescription, as outlined by Ohta [13]. In general, the
form factor is a function of the Lorentz invariants $k^2$, $p^2$ and
$p \cdot k$, where $k$ is the four-momentum of the meson in the loop
and $p$ is the momentum of the initial nucleon. This is, however, not
the case for the static CBM, where we have instead a dependence on
only $k$, with
\begin{equation}
f(k)~=~u(|{\bf k}|R) \ .
\end{equation}

The result of such a calculation [25] is that the e.m. vertex
function of the meson in the loop, $\Gamma^\mu_M$ of Eq. (17),
has to be modified to
\begin{eqnarray}
\tilde\Gamma^\mu_M (q^2)~=~ig^2e_M\int\! {d^4k \over (2\pi)^4}~
\Biggl[~\gamma_5 &  & ~f(k)~S(p-k)~f(k)~\gamma_5~\Delta(k)~- \\
 &  & \gamma_5~f(k')~S(p-k)~f(k')~\gamma_5~\Delta(k')~\Biggr]
{}~{(k+k')^\mu \over (k'^2 - k^2)} \ , \nonumber
\end{eqnarray}
where $e_M$ is the charge of the meson in the loop and where again
pseudoscalar coupling is applied.

Recalling the expressions for the one-loop contribution to the
baryonic self energy, $\Sigma(p)$ of Eq. (12), and $\Gamma^\mu_B$,
the e.m. vertex function that resembles the coupling of the external
photon to the intermediate baryon with charge $e_B$, the second term
in $\Gamma^\mu_F$ of Eq. (20), we can finally verify the following
Ward identities:
\begin{mathletters}
\begin{eqnarray}
q_\mu \cdot \Gamma^\mu_B & = & e_B \left( \Sigma(p) - \Sigma(p')
\right) \ ,  \\
q_\mu \cdot \tilde\Gamma^\mu_M & = & e_M \left( \Sigma(p)-\Sigma(p')
\right) \ ,
\end{eqnarray}
\end{mathletters}
which can be summarized as
\begin{equation}
q_\mu \cdot \Gamma^\mu_B~+~q_\mu \cdot \tilde\Gamma^\mu_M~
=~e_N \left( \Sigma(p) - \Sigma(p') \right)\ ,
\end{equation}
where $e_N$ is the charge of the baryon under investigation. We note,
that the unmodified vertex, $\Gamma^\mu_M$ of Eq. (17), does not
fulfill Eq. (31b). Furthermore, this alteration is necessary also
for other chiral extensions of effective quark soliton models and
for both the surface coupled and the volume coupled version of the
CBM. The latter already has a ``seagull" contribution,
$\Gamma^\mu_{sg}$ of Eq. (26), which however is purely transverse,
i.e. $q_\mu \cdot \Gamma^\mu_{sg}=0$, and hence does not contribute
to the restoration of the Ward-Takahashi identity in Eq. (31b).

This semblant arbitrariness is related to the fact that the
Ward-Takahashi identities restrict only the longitudinal part of the
e.m. current. Hence, they permit the underlying dynamics -- that
gave rise to the form factor and the ``seagull" insertion to begin
with -- to generate additional terms which independently satisfy the
Ward-Takahashi identities [10].

As can be seen from Eq. (29), in the CBM, $f(k)$ does not depend on
$k_0$, the energy of the meson in the loop, as the latter is fixed by
putting the meson on its mass shell. Thus, the time-component of the
``seagull" correction will also vanish and the electric Sachs form
factor, $G_E^{mes}$ of Eq. (18), remains unchanged, while its
magnetic counterpart, $G_M^{mes}$ of Eq. (19), will be modified.
We finally find
\begin{eqnarray}
\tilde G_M^{mes}(|{\bf q}|) = {1\over Z}
\sum_{B} &  & {e_Ms(B)f_{\pi NB}^2 \over 4\pi^2 m_\pi^2} \nonumber \\
&  & \qquad\int\! d^3{\bf k} \left(
{u(|{\bf k'}|R)^2 \over \omega_{\bf k'}(\omega_{\bf k'}+\omega_B)}~-~
{u(|{\bf k}|R)^2  \over \omega_{\bf k}    (\omega_{\bf k}+\omega_B)
}\right) {|{\bf k} \times \hat{\bf q}|^2 \over \omega_{\bf k}^2-
\omega_{\bf k'}^2} \ .
\end{eqnarray}

Another necessary requirement of e.m. gauge invariance is the
conservation of the total electric charge, i.e.
\begin{equation}
G_E(|{\bf q}|=0)~=~e_N\,,
\end{equation}
where $e_N$ is the charge of the nucleon under consideration.  The
CBM fulfills this requirement -- with or without the ``seagull"
insertion -- as can be seen from Eqs. (18) and (21), using
\begin{mathletters}
\begin{eqnarray}
G_E^{mes}(|{\bf q}|=0) & = & {1\over Z}~\sum_{B} e_M Z_B \ , \\
G_E^{bag}(|{\bf q}|=0) &= & {1\over Z}~\left(~e_N~+~\sum_B e_B~Z_B
\right) \ ,
\end{eqnarray}
\end{mathletters}
where $Z=1+\sum_{B} Z_B$ and $e_N=e_M+e_B$.

In Fig. 1, we depict the size of the ``seagull" corrections for a
typical bag radius of $R=1$ fm. The uncorrected magnetic Sachs form
factor of the mesonic cloud, $G_M^{mes}(|{\bf q}|)$ of Eq. (19), is
shown as a dot-dashed line, whereas the gauge invariant corresponding
expression, $\tilde G_M^{mes}(|{\bf q}|)$ of Eq. (33), is shown as a
solid line.  The process under consideration is $p \rightarrow
n+\pi^+ \rightarrow p$ and the normalization factor, $Z^{-1}$, is not
taken into account in this figure.  It is seen that the correction is
significant, e.g.  about 100\% for the magnetic moment, $\mu_M \equiv
G_M^{mes}(|{\bf q}|=0)$.  Also, the dependence on the momentum
transfer is quite different, with the corrected form factor being
much harder.

In Figure 2, we show the magnetic moment, $\mu_M$, as a function
of the MIT bag radius, both for the gauge invariant calculation
(solid line) as well as the original evaluation which violates gauge
invariance (dot-dashed line). Again, the differences are appreciable.
In the limit of very small $R$, the gauge invariance correction
stemming from the ``seagull" terms is almost 50\% of the uncorrected
value and both diverge like $1/R$. In all calculations shown, the
surface coupled version of the CBM was employed.

In a previous publication [23], we investigated the strangeness
content of the nucleon as generated by the kaon cloud, employing, for
example, an $SU(3)$-extended cloudy bag model.  The only free
parameter in the model, the MIT bag radius, was adjusted to give an
optimal description of the standard electromagnetic observables, i.e.
the charge radii and magnetic moments of both the neutron and proton.
Predictions were made for the strange quark contributions to the weak
magnetic form factor, $F_2^s(0)$, the strangeness radius of the
nucleon, $<\!r^2\!>_s$, and the strangeness axial vector
matrix-element, $g_A^s$.  In that work, the ``seagull" diagrams
discussed here were not considered and the strangeness
matrix-elements we found turned out to be quite small.

If we now include gauge invariance corrections, the value we find for
the strange magnetic moment more than triples --  our prediction
changes from $F_2^s(0)=-0.0260$ to $F_2^s(0)=-0.0917$ -- and hence
approaches a magnitude that -- according to ref. [10] -- would be
observable in the corresponding experiments [26,27]. The strangeness
radius and the strange axial vector matrix-element, on the other
hand, are hardly effected by the inclusion of the ``seagull" diagrams
and stay rather small. After restoring the gauge invariance of the
CBM in the fashion outlined here, our findings agree quite well with
those of ref. [10], where analogous strangeness matrix-elements were
modeled by employing kaon loops and effective meson-nucleon vertices
taken from nucleon-nucleon and nucleon-hyperon scattering.

\section{CONCLUSIONS}

By interpreting the cloudy bag model (CBM) as a phenomenological
field theory with free pion and composite nucleon fields coupled via
an effective, momentum dependent vertex, we show, that the
conventional analysis of the e.m. form factors of the nucleon
violates the Ward-Takahashi identities. We then establish the
necessary ``sea\-gull" correction terms, following either the recipe
presented by Gross and Riska [12] or the minimal substitution
prescription, introduced by Ohta [13], and we show that the resulting
e.m.  vertex functions obey the Ward-Takahashi identities. We
demonstrate, that the corrections, which appear only in the magnetic
form factor, arise independently of which version of the CBM --
pseudoscalar (surface) or pseudovector (volume) coupling -- is
employed, and we point out that our findings also apply to other,
chirally invariant hybrid quark models. We then depict the magnitude
of those alterations by comparing the Sachs magnetic form factor and
the magnetic moment with and without those insertions and observe
that they have an appreciable effect. Finally, we discuss the effect
of these alterations on our past investigation of the strangeness
content of the nucleon [23].  Our prediction for the strange magnetic
moment increases dramatically through those corrections and our
results now concur with the findings of ref. [10].

Although the underlying Lagrangian, ${\cal L}$ of Eq. (1), which is
expressed in terms of quark and pion degrees of freedom, is gauge
invariant, the model, when understood in terms of effective,
composite nucleons and elementary pions, violates the Ward-Takahashi
identities. It is, therefore, in the transition from the microscopic
quark degrees of freedom to the effective, composite nucleonic fields
that the $U(1)$ gauge invariance is lost. This issue deserves further
clarification [28] and the prescription outlined here can only serve
as a first step in this direction.

\acknowledgments{We thank Mike Musolf for bringing this issue to our
attention and Steve Pollock, Michael Frank and Matthias Lutz for a
number of fruitful discussions.  This work was supported in part by
the US Department of Energy and by the Deutsche
Forschungsgemeinschaft.}

\begin{figure}
\caption{The contribution of the $\pi^+$ cloud to the magnetic form
factor of the proton (in units of $\mu_N$) evaluated in the
pseudoscalar coupled CBM for a bag radius of $R=1$ fm is shown as a
function of the three-momentum transfer. The solid line depicts the
gauge invariant calculation, Eq. (33), and the dot-dashed line shows
the standard evaluation, Eq. (19), which violates gauge invariance.
The normalization factor of Eq. (16) is not taken into account.}
\end{figure}

\begin{figure}
\caption{The contribution of the $\pi^+$ cloud to the magnetic moment
of the proton (in units of $\mu_N$) evaluated in the pseudoscalar CBM
is depicted as a function of the MIT bag radius. The solid line shows
the gauge invariant calculation and the dot-dashed line shows the
standard evaluation which violates the Ward-Takahashi identities.
Again, the normalization factor of Eq. (16) is not taken into
account.}
\end{figure}

\newpage
\vbox to 4truein{
\centerline{\bf FIG.~1:}
\vskip  3.80truein
\centerline{\epsfbox{f1.ps}\hskip 1.0truein}
\vss}
\vfill
\vbox to 4truein{
\centerline{\bf FIG.~2:}
\vskip  3.80truein
\centerline{\epsfbox{f2.ps}\hskip 1.0truein}
\vss}
\end{document}